\documentclass[11pt,twoside]{article}

\usepackage{asp2006}
\usepackage{epsf}
\usepackage{lscape}

\begin{document}
\title{Unification of 3CR Radio Galaxies and Quasars}   %%% Fill in title
\author{Bernhard~Schulz} 
\affil{California Institute of Technology, IPAC, M/C 100-22, Pasadena, CA 91125, USA}
\author{Ralf~Siebenmorgen}  
\affil{European Southern Observatory, Karl-Schwarzschild Str. 2, D-85748 Garching, Germany }
\author{Martin~Haas} 
\affil{Astronomisches Institut, Ruhr Universit\"at, Universit\"atsstr. 150, D-44780 Bochum, Germany}
\author{Endrik~Kr\"ugel}  
\affil{MPI f\"ur Radioastronomie, Auf dem H\"ugel 69, D-53121 Bonn, Germany}
\author{Rolf~Chini}  
\affil{Astronomisches Institut, Ruhr Universit\"at, Universit\"atsstr. 150, D-44780 Bochum, Germany}

\begin{abstract} %%% Abstract to run on from here.
We have observed seven powerful FR2 radiogalaxies and seven quasars with 
the Spitzer IRS \citep{houck2004}. Both samples are comparable in both, 
redshift range and isotropic 178~Hz luminosity. Both samples are found to
have similar distributions in the luminosity ratios of Mid-IR high- and 
low-excitation lines ([NeV]/[NeII]), and of Mid-IR high-excitation line to 
radio power ratio ([NeV]/$\mathrm{P_{178\: MHz}}$). However, the MIR/FIR ratio 
is generally higher for quasars. We further observed Silicate features at
10 and 18~$\mu$m in emission. 
In our sample only quasars show emission features, while silicate absorption 
is seen only in the radio galaxies. These
observations are all in agreement with unification schemes that explain
both groups as the same class of objects seen under different orientation
angles.
\end{abstract}

According to unification schemes, quasars (Type~1 AGN) and powerful radio 
galaxies (Type~2 AGN) are the same phenomenon seen from different aspect
angles. A dusty molecular torus blocks the line of sight to the central engine
and the broad line region when seen from the side. Narrow high ionization lines
are seen in both AGN types and are thought to originate outside the torus. 
The same is certainly true for the radio emission originating from the lobes of
the radio jets. The radio power is a measure for the total energy output of the
AGN, and is correlated to the FIR luminosity \citep{meise2001}.

The subsamples consist of 7 FR2 radio galaxies (3C079, 3C295, 3C303.1, 3C321, 3C356, 
3C381, 3C459 ) as Type~2 objects and 7 quasars or broad line radio galaxies (3C047, 3C109, 3C249.1, 
3C298, 3C323.1, 3C351, 3C445) as Type~1 objects \citep{haas2005}. 
They are comparable in redshift
($\mathrm{0.05 \leq z \leq 1.5}$) and isotropic 178~MHz luminosity
($\mathrm{10^{26.5}\: W/Hz\leq}$$\mathrm{P_{178\: MHz}}$ $\mathrm{\leq 10^{29.5}\: W/Hz}$). 
The selection was taken  from a subsample of the 3C
catalogue, previously observed with ISO \citep{sieben2004, haas2004} that had
sufficient S/N and was not blocked by Spitzer guaranteed time reservations.

Both samples show similar high to low excitation line ratios ([NeV]/[NeII]) 
and are statistically indistinguishable, as expected for AGN \citep{genzel1998}. 
Both samples show also similar ratios of [NeV] and radio emission. 
The MIR/FIR luminosity, however, is generally higher for quasars 
than for the FR2 radio galaxies. If the FIR luminosity is correlated to the AGN 
power, as suggested by \cite{meise2001}, then the MIR/FIR ratio hints at 
substantial dust absorption in FR2 galaxies, since the radio powers of both 
AGN types in the sample are similar. For a diagram see \cite{haas2005}.

FR2 galaxies show more attenuated visible [OIII]$_{\mathrm{500.7~nm}}$ line emission compared to 
the infrared [OIV]~line, than quasars, probably due to dust absorption. We conclude
therefore that the [OIII]$_{\mathrm{500.7~nm}}$ line is not a good isotropic tracer for
testing unification schemes in high luminosity objects.

The \cite{pierkr1992} dust model for AGN predicted Silicates not only in absorption
but also in emission, as observed with Spitzer IRS \citep{sieben2005, hao2005}.
We see absorption in 3C079, 3C303.1, 3C321, and 3C459, which are all of AGN Type~2,
while emission is observed in 3C109, 3C249.1, 3C323.1, and 3C351, which are all of
Type~1. Within the statistical boundaries of our sample, we find accordingly a good 
correlation of silicate emission with AGN Type~1, and absorption with AGN Type~2.

The $\approx$11~$\mu$m emission bump can be modeled by a simple 3 component 
dust model without postulating exotic grain sizes, abundances, or dust cloud geometries.
We use optically thin emission of Silicate dust with 3 radiating components and a 
primary powerlaw spectrum source radiating at 0.1-15~$\mu$m ($\alpha$=-0.7). The position 
shift of the emission feature towards longer wavelengths is explained by the folding of the 
Silicate absorption coefficient with the steeply rising Planck function. We assumed a 
standard galactic dust mixture of carbon and silicate spheres of 0.1~$\mu$m radius with 
optical constants by \cite{zubko2004} and cross sections calculated from Mie-theory.

\acknowledgements This work is based on observations with the Spitzer Space Telescope, and
has made use of the NASA/IPAC Extragalactic Database (NED), both operated 
by JPL/Caltech, under contract with NASA.

\end{document}